\documentclass[useAMS]{mn2e}
\usepackage{url}
\usepackage{graphicx}
\usepackage{soul}
\usepackage[T1]{fontenc}

\title[]{2015 Southern Taurid fireballs and asteroids 
2005 UR and 2005 TF50}
\author[A. Olech et al.]{A. Olech$^{1}$\thanks{e-mail:
olech@camk.edu.pl},  
P. \.Zo{\l}\k{a}dek$^2$, M. Wi\'sniewski$^{2,3}$, R. Rudawska$^4$, M. B\k{e}ben$^2$, 
\newauthor T. Krzy\.zanowski$^2$, M. Myszkiewicz$^2$, M. Stolarz$^2$, M. Gawro\'nski$^5$, M. Gozdalski$^2$,
\newauthor T. Suchodolski$^6$, W. W\k{e}grzyk$^2$ and Z. Tymi\'nski$^7$\\
$^{1}$Nicolaus Copernicus Astronomical Center,
Polish Academy of Sciences, ul.~Bartycka~18, 00-716~Warszawa, Poland\\
$^{2}$Comets and Meteors Workshop, ul. Bartycka 18, 00-716 Warszawa, Poland\\
$^{3}$ Central Office of Measures, ul. Elektoralna 2, 00-139 Warsaw, Poland\\
$^{4}$ ESA European Space Research and Technology Centre, Noordwijk, The Netherlands\\
$^{5}$ Toru\'n Centre for Astronomy, Faculty of Physics, Astronomy and Applied Informatics, N. Copernicus University,\\  
ul. Grudzi\c{a}dzka 5, 87-100 Toru\'n, Poland\\
$^{6}$ Space Research Centre, Polish Academy of Sciences, ul. Bartycka 18A, 00-716 Warszawa, Poland\\  
$^{7}$ Narodowe Centrum Bada\'n J\k{a}drowych, O\'srodek Radioizotop\'ow POLATOM, 
ul. So{\l}tana 7, 05-400 Otwock, Poland
}

\begin{document}

\date{Accepted 2016 March 15. Received 2016 February 29; in original form 2016 February 1}

\pagerange{\pageref{firstpage}--\pageref{lastpage}} \pubyear{2016}

\maketitle

\label{firstpage}

\begin{abstract}

On the night of Oct 31, 2015 two bright Southern Taurid fireballs
occurred over Poland, being one of the most spectacular bolides of this
shower in recent years. The first fireball - PF311015a Okonek - was
detected by six video stations of Polish Fireball Network (PFN) and 
photographed by several bystanders, allowing for precise determination
of the trajectory and orbit of the event. The PF311015a Okonek entered
Earth's atmosphere with the velocity of $33.2\pm0.1$ km/s and started to
shine at height of $117.88 \pm 0.05$ km. The maximum  brightness of
$-16.0 \pm 0.4$ mag was reached at height of $82.5\pm0.1$ km. The
trajectory of the fireball ended at height of $60.2\pm0.2$ km with
terminal velocity of $30.2\pm1.0$ km/s.

The second fireball - PF311015b Ostrowite - was detected by six video
stations of PFN. It started with velocity of $33.2\pm0.1$ km/s at
height of $108.05 \pm 0.02$ km. The peak brightness of $-14.8 \pm 0.5$ mag
was recorded at height of $82.2\pm0.1$ km. The terminal velocity was
$31.8\pm0.5$ km/s and was observed at height of $57.86\pm0.03$ km.

The orbits of both fireballs are similar not only to orbits of Southern
Taurids and comet 2P/Encke, but even closer resemblance was noticed for
orbits of 2005 UR and 2005 TF50 asteroids. Especially the former
object is interesting because of its close flyby during spectacular
Taurid maximum in 2005. We carried out a further search to investigate
the possible genetic relationship of Okonek and Ostrowite fireballs with
both asteroids, that are considered to be associated with Taurid
complex. Although, we could not have confirmed unequivocally the
relation between fireballs and these objects, we showed that both
asteroids could be associated, having the same origin in a disruption
process that separates them.

\end{abstract}


\section{Introduction}

The Taurids are an annual meteor shower active in October and November
with maximum Zenithal Hourly Rates of about 5. The orbit of the shower
has low inclination and thus, due to the gravitational perturbations of
planets, swarm of particles is diffuse and separated into two main
branches i.e. Northern Taurids (NTA) and Southern Taurids (STA). 

The parent body of Taurid complex is comet 2P/Encke (Whipple 1940),
however both 2P/Encke and Taurids are believed to be remnants of a much
larger object, which has disintegrated over the past 20000 to 30000
years (Asher et al. 1993, Babadzhanov et al. 2008). Recently,
Porub\v{c}an et al. (2006) identified as many as 15 Taurid complex
filaments and found possible association with 9 Near Earth Objects
(NEOs). Most recently, Jopek (2011) identified as many as 14 parent
bodies of the Taurids stream. 

It has been widely recognized that the Taurid complex, despite its
moderate activity, produces a great number of bright fireballs. Asher
(1991) suggested that a swarm of Taurids being in 7:2 resonance with
Jupiter  produces occasional enhanced activity. It was later confirmed
by Asher and Izumi (1998) who predicted observed swarm encounters in 1998,
2005 and 2008. In period 2009-2014 activity of Taurid shower was typical with
some fireballs were observed (Madiedo et al. 2011, 2014).
The return in 2005 was spectacular with both enhanced
global activity and maximum rich in fireballs (Dubietis and Arlt, 2006).
In 2008 the activity of the shower was lower but still it may be
considered as enhanced (Jenniskens et al. 2008, Shrben\'y and Spurn\'y
2012). According to the Asher's model, the next swarm encounter year was
expected in 2015.

Additionally,  most recently, the Taurid shower was suspected to have
ability to produce meteorites (Brown et al. 2013, Madiedo et al. 2014).
On the other hand, Tubiana et al. (2015) found no spectroscopic evidence
for a link between 2P/Encke, the Taurid complex NEOs and CM type
carbonaceous chondrite  meteorites which felt recently in Denmark and
were suspected for origin from the Taurid-Encke complex. Moreover, there
is still no consensus concerning origin of the complex while the
spectral data of its largest objects do not support a common cometary
origin (Popescu et al. 2014).

In this paper we report the results of observations and data reduction
of two very bright Taurid fireballs which were detected on 2015 October
31 over Poland. Afterwards we discuss their connection with the
asteroids 2005 UR and 2005 TF50, and comet 2P/Encke. This is the pilot
study preceding more comprehensive analysis devoted to an enhanced
fireball activity of 2015 Taurid meteor complex and its comparison to
2005 Taurid return.  

\section{Observations}

The PFN is the project whose main goal is regularly monitoring the sky
over Poland in order to detect bright fireballs occurring over the whole
territory of the country (Olech et al. 2006, \.Zo{\l}\k{a}dek et al.
2007, 2009, Wi\'sniewski et al. 2012). It is kept by amateur astronomers
associated in Comets and Meteors Workshop (CMW) and coordinated by
astronomers from Copernicus Astronomical Center in Warsaw, Poland.
Currently, there are almost 30 fireball stations belonging to PFN that
operate during each clear night. It total over 60 sensitive CCTV cameras
with fast and wide angle lenses are used. 

\begin{figure}
\centering
\includegraphics{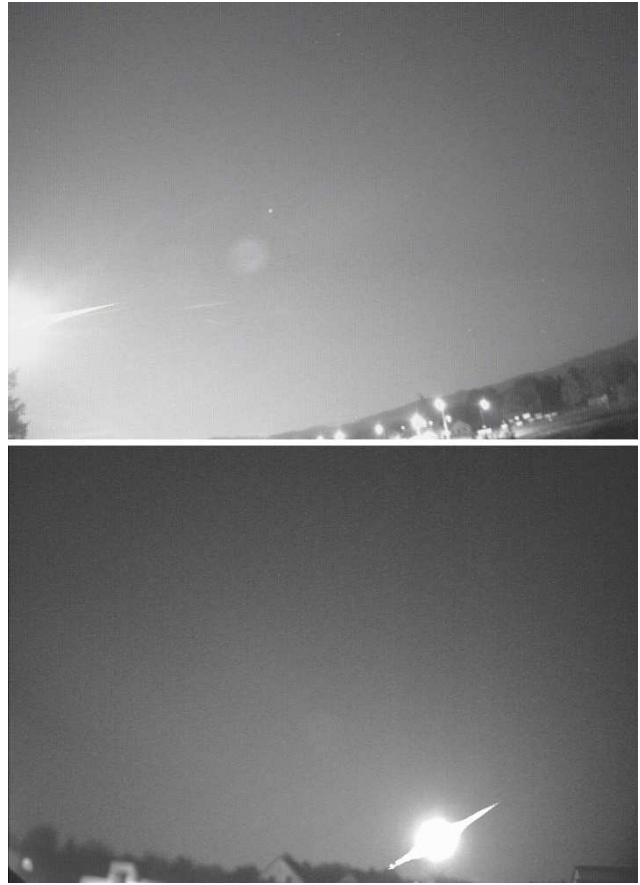}  
\vspace{12cm}
\caption{The video images of PF311015a Okonek fireball captured in
Podg\'orzyn (upper panel) and Rzesz\'ow (lower panel).}
\end{figure}

\begin{figure}
\centering
\includegraphics{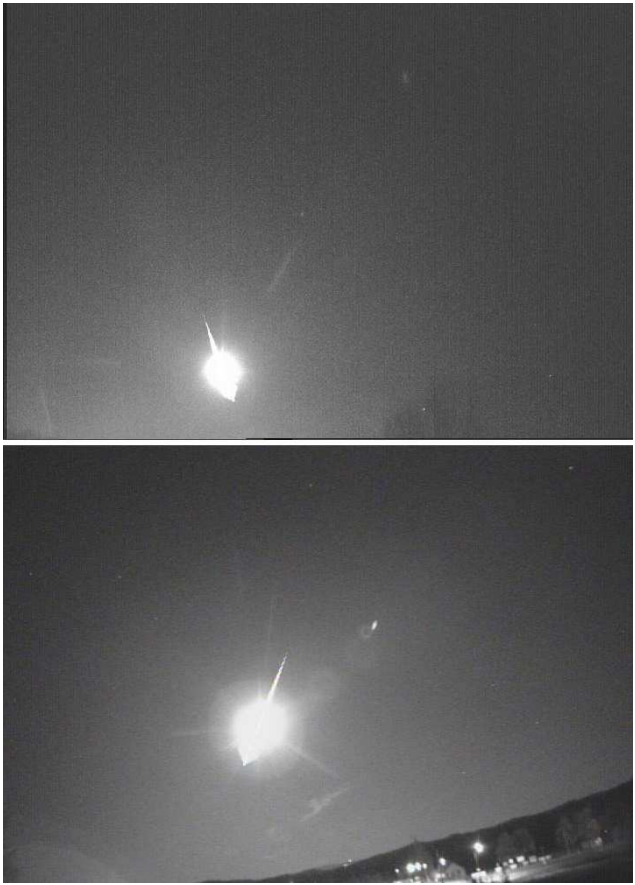}   
\vspace{12cm}
\caption{The video images of PF311015b Ostrowite fireball captured in
Urz\k{e}d\'ow (upper panel) and Podg\'orzyn (lower panel).}
\end{figure}

\begin{table*}
\centering
\caption[]{Basic data on the PFN stations which recorded PF311015a Okonek fireball.}
\begin{tabular}{|l|l|c|c|c|l|l|}
\hline
\hline
Code & Site & Longitude [$^\circ$] & Latitude [$^\circ$] & Elev. [m] & Camera & Lens \\
\hline
PFN38 & Podg\'orzyn & 15.6817 E & 50.8328 N & 360 & Tayama C3102-01A4 & Computar 4mm f/1.2 \\
PFN38 & Podg\'orzyn & 15.6817 E & 50.8328 N & 360 & KPF 131HR & Panasonic 4.5mm f/0.75 \\
PFN43 & Siedlce & 22.2833 E & 52.2015 N & 152 & Mintron MTV-23X11C & Ernitec 4mm f/1.2 \\
PFN48 & Rzesz\'ow & 21.9220 E & 50.0451 N & 230 & Tayama C3102-01A4 & Computar 4mm f/1.2 \\
PFN52 & Stary Sielc & 21.2923 E & 52.7914 N & 90 & DMK23GX236 & Tamron 2.4-6mm f/1.2 \\
PFN61 & Piwnice & 18.5603 E & 53.0951 N & 85 & Tayama C3102-01A4 & Ernitec 4mm f/1.2 \\
PFN67 & Nieznaszyn & 18.1849 E & 50.2373 N & 200 & Mintron MTV-23 X11E & Panasonic 4.5mm f/0.75 \\
\hline
\hline
\end{tabular}
\end{table*}

During last ten years typical setup of the PFN station consisted of 2-3
Tayama C3102-01A4 cameras equipped with 4 mm f/1.2 Computar or Ernitec
lenses. Tayama C3102-01A4 is cheap CCTV camera with 1/3" Sony SuperHAD
CCD detector working in PAL interlaced resolution with 25 frames per    
second. The field of view of one camera with 4 mm lens  is
$69.8\times55.0$ deg with scale of $\sim 10'$/pixel. This setup allows
detection of the atmospheric entries of debris (both natural and
artificial) with accuracy of trajectory determination below 300 meters.

Almost each station is equipped with a PC computer with Matrox Meteor II
frame grabber. The signal from each camera is analyzed on-line. Our
video stations use the {\sc MetRec} software (Molau 1999) which automatically
detects meteors in frames captured by Matrox Meteor II frame grabber.
The frames containing meteors are stored into BMP files. Additionally,
information about basic parameters of the event such as its time of
appearance and $(x,y)$ coordinates on the frame are saved into {\sc MetRec}
INF files. In case of some new stations containing higher resolution
cameras, {\sc UFO Capture} software is used (SonotaCo 2009). 

On the evening of 2015 October 31 at 18:05 UT a very bright fireball
appeared over northwestern Poland. The International Meteor Organization
(IMO) received almost 70 visual reports concerning this event from
Austria, Czech Republic, Denmark, Germany, Netherlands, Poland and
Sweden\footnote{http://imo.net/node/1645}. Good weather conditions that
night in central Europe and the high brightness of the fireball were the
main factors for receiving so many reports. However, there are also two
other reasons for it as well.

Firstly, the fireball appeared almost exactly at the moment of the close
flyby of large 2015 TB145 asteroid. Such events attract not only the
attention of astronomy amateurs but also general public, encouraging
people for observations. What is more interesting, the asteroid was
passing across the Ursa Major constellation at that time, which was
exactly the same region of the sky where the fireball was visible for
observers situated in central and southern Poland, Czech Republic and
Slovakia. Finally, the date of the appearance of the fireball was the
date of the All Saints' Eve. This is a public holiday in Poland, when in
the evening people gather in  cemeteries lighting the candles at the
graves of their relatives. The scenery of candles after dusk creates
nice landscapes which many people try to photograph. It is thus no
surprise then that one of the most beautiful images of this fireball was
captures in the cementary in Czernice Borowe, Poland by Aleksander and
Grzegorz Zieleniecki. This image, kindly shared by the authors, was used
for analysis in this work.

The fireball reached its maximum brightness over Okonek city, and
therefore received designation PF311015a Okonek. It was observed by six
regular PFN video stations, where from the PFN38 Podg\'orzyn station the
fireball was recorded by two cameras.  Basic properties of the stations
that recorded the event are listed in Table 1. Figure 1 shows images of
the fireball captured by the Podg\'orzyn and Rzesz\'ow stations.
Additionally, the bolide was accidentally photographed in Czernice Borowe
(location of the former PFN22 station). Here the Nikon D3300 digital
single-lens reflex camera with Nikkor AF-S 18-55 mm f/3.5-5.6 lens was
used. The lens was set at 18 mm focal length with relative aperture of
f/3.5. The exposure time was 20 seconds with ISO equal to 800.

Five hours after the PF311015a Okonek appearance another very bright
fireball appeared passed through the sky at 23:13 UT. It was only
slightly fainter than Okonek fireball but due to the late hour (after
midnight of local time) it was not observed by any bystanders.
Fortunately, it was detected by six PFN stations which are listed in
Table 2 and the images of the fireball captured by the Urz\k{e}d\'ow and
Podg\'orzyn stations are shown in Figure 2. The maximum brightness of
the meteor was observed over Ostrowite village and thus its designation
is PF311015b Ostrowite.

\begin{table*}
\centering   
\caption[]{Basic data on the PFN stations which recorded PF311015b Ostrowite fireball.}
\begin{tabular}{|l|l|c|c|c|l|l|}
\hline
\hline
Code & Site & Longitude [$^\circ$] & Latitude [$^\circ$] & Elev. [m] & Camera & Lens \\
\hline
PFN20 & Urz\k{e}d\'ow & 22.1456 E & 50.9947 N & 210 & Tayama C3102-01A1 & Ernitec 4mm f/1.2 \\
PFN38 & Podg\'orzyn & 15.6817 E & 50.8328 N & 360 & Tayama C3102-01A4 & Computar 4mm f/1.2 \\
PFN38 & Podg\'orzyn & 15.6817 E & 50.8328 N & 360 & KPF 131HR & Panasonic 4.5mm f/0.75 \\
PFN43 & Siedlce & 22.2833 E & 52.2015 N & 152 & Mintron MTV-23X11C & Ernitec 4mm f/1.2 \\
PFN48 & Rzesz\'ow & 21.9220 E & 50.0451 N & 230 & Tayama C3102-01A4 & Computar 4mm f/1.2 \\
PFN52 & Stary Sielc & 21.2923 E & 52.7914 N & 90 & Watec 902B & Computar 2.6mm f/1.0 \\
PFN57 & Krotoszyn & 17.4416 E & 51.7018 N & 150 & Tayama C3102-01A4 & Computar 4mm f/1.2 \\
\hline
\hline
\end{tabular}
\end{table*}

All analog video cameras contributing to this paper work in PAL
interlaced resolution of $768\times 576$ pixels, with 25 frames per sec 
offering 0.04 sec temporal resolution. While, the digital camera DMK23GX236
used in PFN52 station has resolution of $1920\times 1200$ pixels and 
works with 20 frames per sec.

\section{Data reduction}

The data from all stations, after a previous conversion, were further
reduced astrometricaly by the {\sc UFO Analyzer} program (SonotaCo
2009). Initially only automatic data were taken into account. However,
during the further processing it became obvious that significant
overexposures, the presence of the wake and a possible fragmentation
after the flare caused quite serious errors concerning the correct
position of the points of the phenomenon. The measurement precision
improved noticeably when the bolide's position was determined using UFO
Analyzer astrometric solution with  manual centroid measurement {\sc UFO
Analyzer}.

The trajectory and orbit of both fireballs was computed using {\sc PyFN}
software (\.Zo{\l}\k{a}dek 2012). {\sc PyFN} is written in Python with
usage of SciPy module and CSPICE library. For the purpose of trajectory
and orbit computation it uses the plane intersection method described by
Ceplecha (1987). Moreover, {\sc PyFN} accepts data in both {\sc MetRec}
(Molau 1999) and {\sc UFOAnalyzer} (SonotaCo 2009) formats and allows
for semi-automatic search for double-station meteors.

In case of photographic image recorded in Czernice Borowe the 
astrometry was performed using {\sc Astro Record 3.0} software (de
Lignie 1997), with the accuracy of the meteor path determination of 3
arcmin. The image was not used for brightness estimate due to the fact
that the meteor was visible through thin cirrus clouds.

\section{Results}

\subsection{Brightness determination of the fireballs}

The photometry of both fireballs was not trivial because of the strong
saturation observed. On every camera both fireballs looks like strongly
overexposed bulbs of light. Video cameras in the northernmost stations
were completly overexposed with whole image saturated. Only the most
distant stations can be usable to any kind of photometric measurements.
The best results has been obtained using PFN48 Rzesz\'ow video
recordings. Both fireballs were recorded completly and from the large
distance. The Okonek fireball has been observed from the distance
of 525 km (point of maximum brightness), the Ostrowite fireball from the
distance of 380 km. From such large distance both fireballs appeared as
overexposed objects with only slightly different brigtness. The
same camera recorded also the Full Moon image and its brightness 
was used as a primary reference point.

\begin{figure}
\centering
\includegraphics{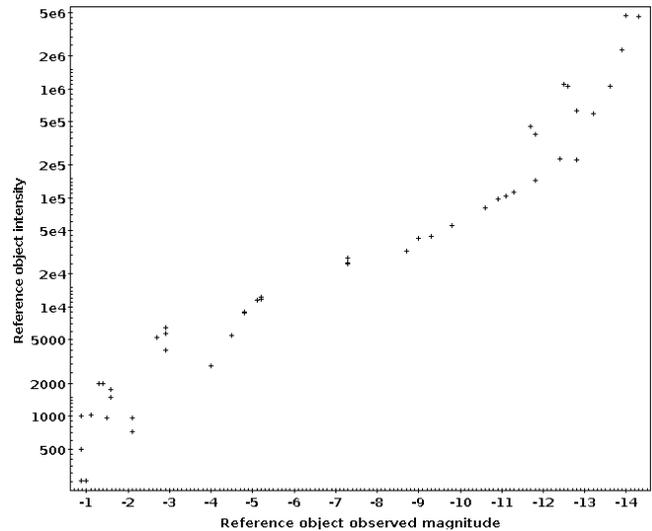}
\vspace{7.7cm}
\caption{Measured intensities and observed magnitudes of the reference objects (Y 
axis logarithmic).}
\end{figure} 

Two independent methods has been applied to estimate the real brightness
of fireballs. The first one was based on the measurements of the sky
brightness during the fireball flight. These measurements has been
compared with the sky brigtness caused by the Full Moon in the same
camera. As a result we had observed maximum magnitude for Okonek
fireball reaching $-12.5$  and $-12.3$ magnitude for the second
fireball. Due to low sensitivity of the cameras used and severe light
pollution in the PFN48 site the resulted lightcurve is incomplete and
contains only brightest part of the fireball. The second method used
several comparison objects recorded by cameras of different types with
the same optics as used in PFN48 and working in similar sky conditions.
The only available comparison objects were bright planets like Jupiter
and Venus (magnitudes in the range $-2.5$ to $-4.5$) and the Moon in the
different phases (magnitudes from $-8$ do $-12.6$). We used two sets of
reference objects - one set recorded by the same camera configuration as
for both fireballs and one set recorded by camera with two magnitude
higher sensitivity. This second set can be treated as a set of reference
objects which are brighter by two magnitude and it is helpfull to fill
the gap between -4.5 and -8 magnitude objects. Some trial and error
photometric tests led us to choose the best measure method. Good results
has been obtained using aperture photometry on the images with
overexposed pixel only visible (pixel with value below 255 on eight bit
image has been rejected). Further refinements lowered reject value to
230. From such measurements the $I(m)$ function has been derived (see
Figure 3). It is an exponential function empiricaly defined in the form:

\begin{equation}
I = A^{-0.45\cdot m}
\end{equation}

\noindent where $I$ is intensity, $m$ magnitude and $A$ constant.

In case of our meaurements this function is valid for magnitude
range from $-4$ to $-12$ and is a bit different for higher magnitudes. 

This method has been used for both fireballs. The light curves has been
determined. Measurements has been repeated using different sets of
reference points, different results has been used to magnitude error
determination. Maximum magnitudes measured using this method were
$-12.4$ for Okonek fireball and $-11.9$ for Ostrowite fireball,
respectively. These results are consistent with measuremets of the sky
brightness mentioned before. Difference between two methods is less than
0.5 mag. Both fireballs observed from PFN48 Rzesz\'ow, from the distance
of hundreds kilometers, looked as very bright objects with brightness
comparable to the Full Moon. Reduction to the standard absolute
magnitude (fireball visible 100 km directly overhead) gives
significantly higher brightness values. Resulting absolute brightness
for Okonek fireball is $-16\pm0.4$ mag. The Ostrowite fireball was
fainter and its absolute magnitude was $-14.8\pm0.5$. Both fireballs
iluminated the southern parts of the country comparable to the Full
Moon. In the north-western part of Poland these fireballs were observed as
extremely bright objects which lit the sky with bright blue-greenish
light.

\subsection{Observational properties of the fireballs}

The PF311015a Okonek fireball appeared over Western Pomerania moving
almost directly from east to west. The beginning of the meteor was
recorded 117.88 km over the Radodzierz Lake. The entry velocity was
$33.2 \pm 0.1$ km/s and was slightly higher than the mean velocity of
Southern Taurids of 28 km/s\footnote{Meteor Data Center:\\ 
http://www.astro.amu.edu.pl/$\sim$jopek/MDC2007/}. The peak brightness was observed
at the height 82.5 km about 10 km east over Okonek city. The fireball
travelled its 181.2 km luminous path in 5.62 seconds, ending at the height
of 60.2 km over Z{\l}ocieniec. The basic characteristics of the PF311015a
Okonek fireball are summarized in Table 3 and its luminous trajectory is
shown in Figure 4.

\begin{figure}
\centering
\includegraphics{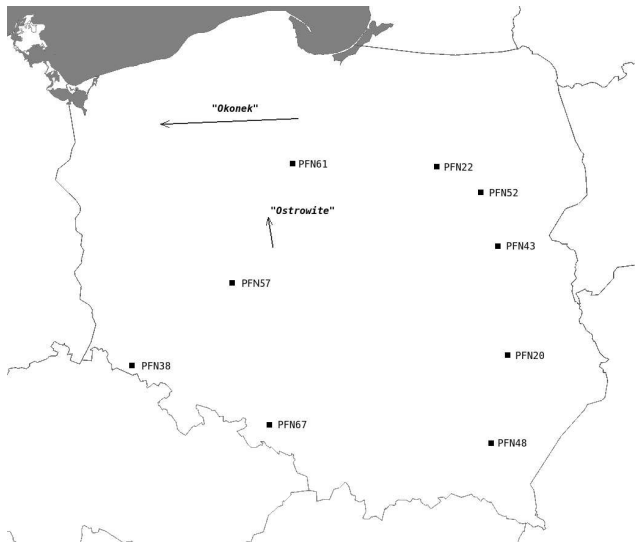}
\vspace{7.8cm}
\caption{The luminous trajectory of the PF311015a Okonek and PF31015b Ostrowite fireballs over Poland 
and the location of the PFN stations which data were used in calculations.}
\end{figure}  

\begin{table*}
\centering
\caption{Characteristics of the PF31015a Okonek fireball}
\begin{tabular}{lccc}
\hline
\multicolumn{4}{c}{2015 October 31, ${\rm T} = 18^h05^m14^s \pm 1.0^s$ UT}\\
\hline
\multicolumn{4}{c}{Atmospheric trajectory data}\\
\hline
 & {\bf Beginning} & {\bf Max. light} & {\bf Terminal} \\
Vel. [km/s] & $33.2\pm0.1$ & $33.0\pm1.0$ & $30.2\pm1.0$ \\
Height [km] & $117.88\pm0.05$ & $82.5\pm0.1$ & $60.2\pm0.2$ \\
Long. [$^\circ$E] & $18.602\pm0.001$ & $17.04\pm0.01$ & $16.020\pm0.002$\\
Lat. [$^\circ$N] & $53.6292\pm0.0004$ & $53.57\pm0.01$ & $53.526\pm0.001$\\
Abs. magn. & $-0.8\pm0.5$ & $-16.0\pm0.4$ & $-2.3\pm0.3$ \\
Slope [$^\circ$] & $19.31\pm0.05$ & $18.39\pm0.05$ & $17.78\pm0.05$\\
Duration & \multicolumn{3}{c}{5.62 sec}\\
Length & \multicolumn{3}{c}{$181.2\pm0.2$ km}\\
Stations & \multicolumn{3}{c}{Podg\'orzyn, Siedlce, Rzesz\'ow, Stary Sielc, Piwnice, Nieznaszyn}\\
\hline
\multicolumn{4}{c}{Radiant data (J2000.0)}\\
\hline
 & {\bf Observed} & {\bf Geocentric} & {\bf Heliocentric} \\
RA [$^\circ$] & $50.10\pm0.08$ & $51.06\pm0.07$ & - \\
Decl. [$^\circ$] & $17.10\pm0.06$ & $15.11\pm0.07$ & - \\
Vel. [km/s] & $33.2\pm0.1$ & $31.0\pm0.1$ & $37.2\pm0.1$\\
\hline
\end{tabular}
\end{table*}

The PF311015a Okonek fireball appeared as a meteor with absolute
magnitude of $-0.8\pm0.5$. The brightness was increasing slowly by about
2 mag throughout the first second of the flight. While during the next
1.5 seconds much more steep increase of brightness was observed, ending
with plateau lasting about one second when the peak brightness reaching
$-16.0\pm0.4$ mag was recorded. The plateau finished abruptly at around
3.8 second of the flight with steep and almost linear decrease of
brightness. At the terminal point the brightness of the meteor was
$-2.3\pm0.5$ mag. At the end of the plateau phase a bright persistent
train appeared. It started to shine with absolute magnitude of $-15$,
and it slowly faded to $-13$ mag during almost two seconds. After that
moment its brightness started to decrease much faster. Figure 5 shows
the light curve of the PF311015a Okonek fireball and its persistent
train.

\begin{figure}
\centering
\includegraphics{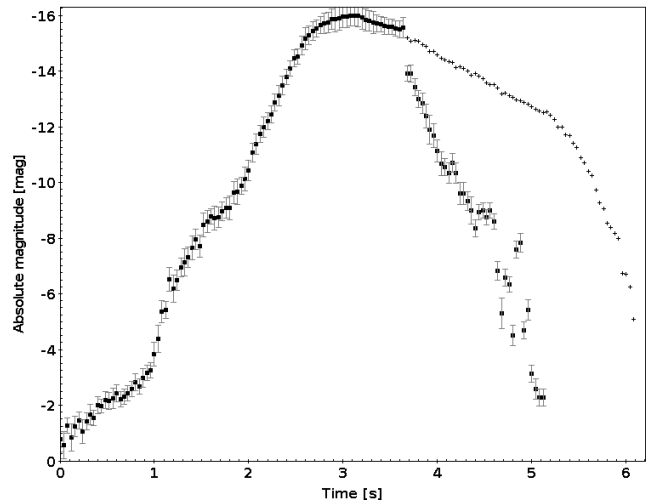}
\vspace{7.2cm}
\caption{The light curve of the PF311015a Okonek fireball (squares) and 
its persistent train (crosses).}
\end{figure}

The PF311015b Ostrowite fireball appeared over the central Poland moving
almost from south to north. It started its luminous path at the height
of 108.05 km over place located about 5 km west of Konin. The entry
velocity was $33.2 \pm 0.1$ km/s and was identical to the velocity of
Okonek fireball. The peak brightness was observed at height of 82.2 km
over the place located 5 km east of Ostrowite. The fireball travelled
its 62.77 km luminous path in 2.0 seconds and ended at height 57.86 km
with terminal velocity equal to $31.8 \pm 0.5$ km/s. The basic
characteristics of the PF311015b Ostrowite fireball are summarized in
Table 4, while its luminous trajectory is shown in Figure 4.

\begin{table*}
\centering   
\caption{Characteristics of the PF31015b Ostrowite fireball}
\begin{tabular}{lccc}
\hline
\multicolumn{4}{c}{2015 October 31, ${\rm T} = 23^h13^m00^s \pm 1.0^s$ UT}\\
\hline
\multicolumn{4}{c}{Atmospheric trajectory data}\\  
\hline
 & {\bf Beginning} & {\bf Max. light} & {\bf Terminal} \\
Vel. [km/s] & $33.2\pm0.1$ & $32.2\pm0.5$ & $31.8\pm0.5$ \\
Height [km] & $108.05\pm0.02$ & $82.2\pm0.1$ & $57.86\pm0.03$ \\
Long. [$^\circ$E] & $18.1654\pm0.0003$ & $18.11\pm0.01$ & $18.0685\pm0.0007$\\
Lat. [$^\circ$N] & $52.2005\pm0.0002$ & $52.37\pm0.01$ & $52.5313\pm0.0004$\\
Abs. magn. & $-0.3\pm1.2$ & $-14.8\pm0.5$ & $-3.6\pm0.4$ \\
Slope [$^\circ$] & $52.29\pm0.05$ & $53.12\pm0.05$ & $52.96\pm0.05$\\
Duration & \multicolumn{3}{c}{2.0 sec}\\
Length & \multicolumn{3}{c}{$62.77\pm0.2$ km}\\
Stations & \multicolumn{3}{c}{Urz\k{e}d\'ow, Podg\'orzyn, Siedlce, Rzesz\'ow}\\
\hline
\multicolumn{4}{c}{Radiant data (J2000.0)}\\
\hline
 & {\bf Observed} & {\bf Geocentric} & {\bf Heliocentric} \\
RA [$^\circ$] & $52.09\pm0.06$ & $51.45\pm0.06$ & - \\
Decl. [$^\circ$] & $15.78\pm0.07$ & $14.61\pm0.06$ & - \\
Vel. [km/s] & $33.2\pm0.1$ & $31.23\pm0.109$ & $37.37\pm0.06$\\
\hline
\end{tabular}
\end{table*}

\begin{figure}
\centering
\includegraphics{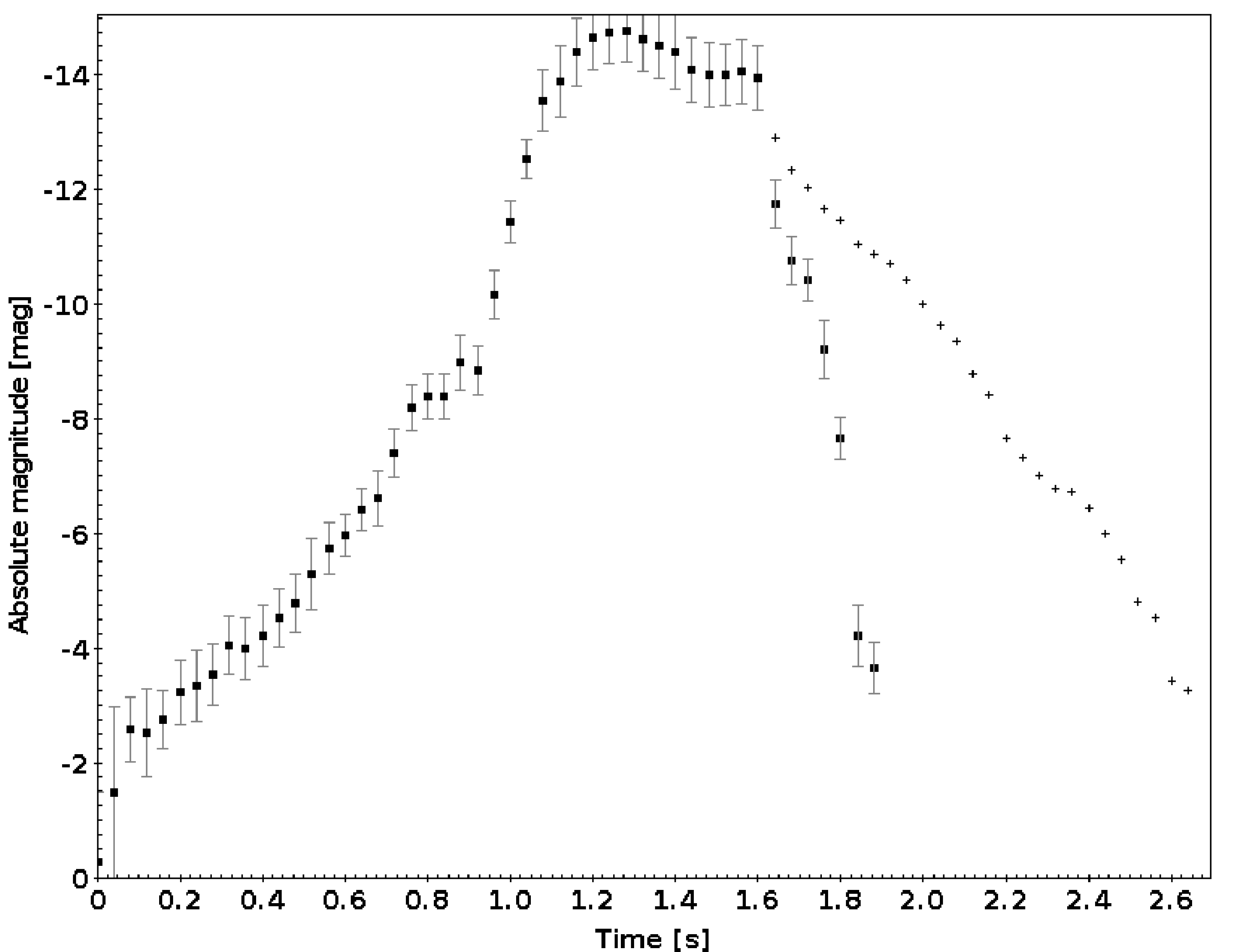}
\vspace{7.2cm}
\caption{The light curve of the PF311015b Ostrowite fireball (squares) and
its persistent train (crosses).}
\end{figure}

The PF311015b Ostrowite fireball appeared as object with absolute
magnitude of $-0.3\pm1.2$. During the first second of the flight its
brightness was increasing almost linearly to the value of $-9$ mag.
After that, within a period of about 0.2 seconds, the brightness of the
fireball increased by a factor of 100, reaching the plateau phase
lasting for about 0.5 seconds. During this plateau the maximum absolute
magnitude of $-14.8\pm0.5$ was recorded. While starting from 1.7 second
of the flight the sudden drop of brightness was observed. So that the
luminous path ended after about 2 seconds of the flight, with he terminal
magnitude equal to $-3.6\pm0.4$.

As in case of Okonek fireball, the bright persistent train was  observed
as well. It started to be visible just after the plateau phase with
brightness of $-13$ mag, and within over one second it faded to $-3$ 
magnitude. The light curve of the PF311015b Ostrowite fireball and its
persistent train is plotted in Figure 6.

\subsection{Orbits of the fireballs}

The orbital parameters of the Okonek and Ostrowite fireballs as computed
from our observations are shown in Table 5. For comparison the orbital
parameters of comet 2P/Encke are listed as well. The orbits of both
fireballs are located almost in the ecliptic plane and have high
eccentricity with perihelion distance slightly less than 0.3 a.u.
The similarity of both orbits is evident with the Drummond criterion
$D_D$ (Drummond 1981), and is equal to only 0.011.

\begin{table*}
\caption[]{Orbital elements of the PF311015a Okonek and PF311015b Ostrowite
fireballs compared to the orbits of
2005 UR and 2005 TF50 near Earth asteroids and comet 2P/Encke}
\centering
\begin{tabular}{|l|c|c|c|c|c|c|c|}
\hline
\hline
  & $1/a$ & $e$ & $q$ & $\omega$ & $\Omega$ & $i$ & P \\
  & [1/AU] & & [AU] & [deg] & [deg] & [deg] & [years] \\ 
\hline
PF311015a Okonek & 0.4440(55) & 0.8690(21) & 0.2948(17) & 120.8(2)   & 37.77623(1) & 4.73(12)   & 3.379(73) \\
PF311015b Ostrowite & 0.4408(50) & 0.8720(17) & 0.2903(15) & 121.2(2)   & 37.98982(1) & 5.636(51)  & 3.51(7) \\
2005 UR   & 0.4420(36) & 0.8797(15) & 0.2723(1)  & 141.03(14) & 19.555(147) & 6.972(26)  & 3.40(4) \\
2005 TF50 & 0.4401(5)  & 0.8689(3)  & 0.2978(3)  & 159.898(8) &  0.666(23)  & 10.699(14) & 3.425(6) \\
2P/Encke  & 0.45144(1) & 0.84833(1) & 0.33596(1) & 186.546(1) & 334.5682(1) & 11.7815(1) & 3.29698(1) \\
\hline
\hline
\end{tabular}
\end{table*}  

\begin{table*}
\caption[]{Drummond criterion $D_D$ values for PF311015a Okonek and 
PF311015b Ostrowite fireballs, 2005 UR and 2005 TF50 
near Earth asteroids and comet 2P/Encke.}
\centering
\begin{tabular}{|l|c|c|c|c|c|}
\hline
\hline
Object    & 2P/Encke & 2005 UR & 2005 TF50 & PF311015a Okonek & PF 311015b Ostrowite \\
 \hline
2P/Encke  &   -   & 0.119 & 0.072 & 0.093 & 0.099 \\
2005 UR   & 0.119 &   -   & 0.052 & 0.044 & 0.036 \\
2005 TF50 & 0.072 & 0.052 &   -   & 0.045 & 0.042 \\
PF311015a Okonek    & 0.093 & 0.044 & 0.045 &   -   & 0.011 \\
PF311015b Ostrowite & 0.099 & 0.036 & 0.042 & 0.011 &   -   \\
\hline
\hline
\end{tabular}
\end{table*}

Comparison of the orbits of Okonek and Ostrowite fireballs to orbits of
asteroids listed in Near Earth Objects - Dynamic Site
(NEODyS-2)\footnote{http://newton.dm.unipi.it/neodys/index.php} allowed
us to select couple of asteroids with Drummond criterion $D_D<0.109$. Two
of them are especially interesting.

The Apollo type 2005 TF50 asteroid was discovered on 2005 October 10 by
M. Block at the Steward Observatory, Kitt Peak. The absolute magnitude
of the object is 20.3 mag which indicates the size of 260-590 meters.
Its Tisserand parameter has value of 2.933. 2005 TF50 was listed in
Porub\v{c}an et al. (2006) as one of NEOs connected with 2P/Encke and Taurid
complex meteor showers. It is close to 7:2 resonance with Jupiter - the
same resonance that was suggested as a possible source of Taurid
enhanced activity by Asher (1991). The Drummond criterion describing the
similarity of the orbit of 2005 TF50 to orbits of Okonek and Ostrowite
fireballs is 0.045 and 0.042, respectively.

The 2005 UR asteroid was discovered by Catalina Sky Survey on 2005
October 23. It belongs to the Apollo group and has a Tisserand parameter
of 2.924. The absolute magnitude of the object is 21.6 mag which
indicates the size of 140-320 meters. 2005 UR was not included in work
of Porub\v{c}an et al. (2006) most probably due to the fact that paper
was already written when the asteroid was discovered. On the other hand
it was listed by Jopek (2011) as one of the parent bodies of the Taurid
complex. The similarity of orbits of Okonek and Ostrowite fireballs to
the orbit of 2005 UR is even more evident than in case of 2005 TF50 with
Drummond criterion values 0.044 and 0.036, respectively.

The interesting fact about 2005 UR asteroid is its close approach to
Earth on 2005 October 30 at 13:11 UT with the distance of 0.041 a.u. The
time of the close passage is at exactly the same moment as outburst of
fireball activity of 2005 Taurids (Dubietis and Arlt, 2006).

\begin{figure*}
\centering
\includegraphics{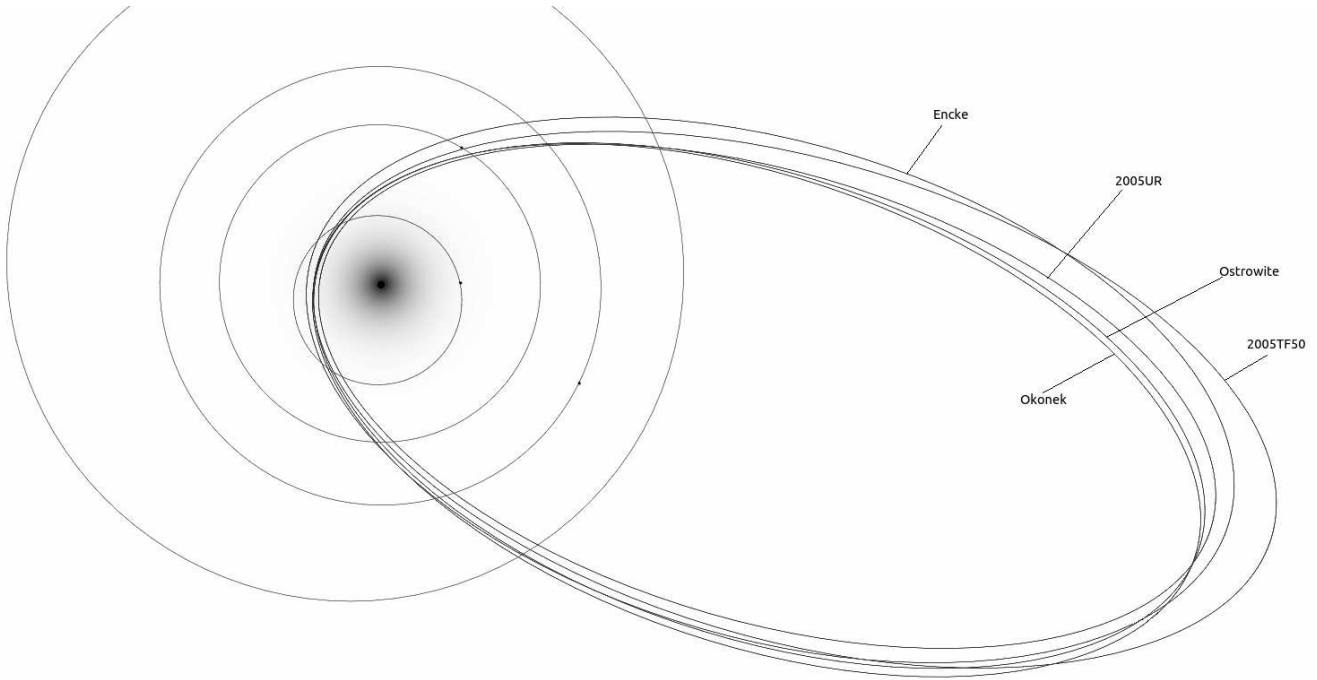}
\vspace{9.5cm}
\caption{Plot of the inner Solar System with orbits of the 
PF311015a Okonek and PF311015b Ostrowite fireballs,
two asteroids: 2005 UR and 2005 TF50 and comet 2P/Encke.}
\end{figure*}

Table 6 lists the Drummond criterion $D_D$ values for PF311015a Okonek
and PF311015b Ostrowite fireballs, 2005 UR and 2005 TF50 Near Earth
Asteroids and comet 2P/Encke. Additionally, Figure 7 shows orbits
of both fireballs in the inner Solar System together with the orbits
of 2005 UR and 2005 TF50 asteroids and comet 2P/Encke.

\begin{figure*}
\centering
\includegraphics{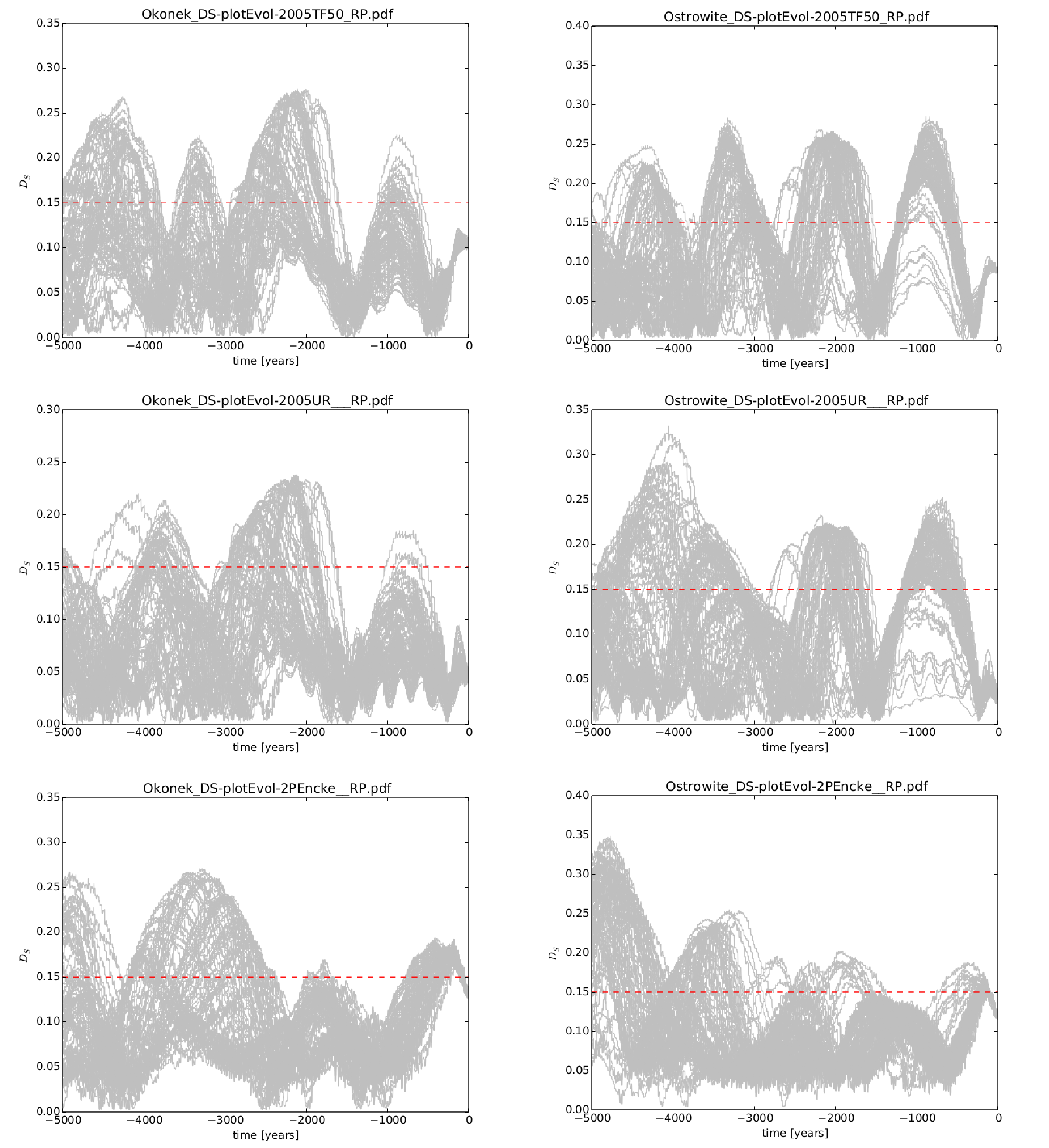}
\vspace{21cm}
\caption{Left panels: evolution of the $D_S$ criterion calculated by comparing the orbit of 2005~TF50, 2005~UR, 2P/Encke
and test particles of PF311015a Okonek fireball. The red dashed line shows the threshold value ($D_c=$0.15).
Right panels: evolution of the $D_S$ criterion calculated by comparing the orbit of 2005~TF50, 2005~UR, 2P/Encke
and test particles of PF31102015b Ostrowite fireball. The red dashed line shows the threshold value ($D_c=$0.15)}
\end{figure*}

\begin{figure*}
\centering
\includegraphics{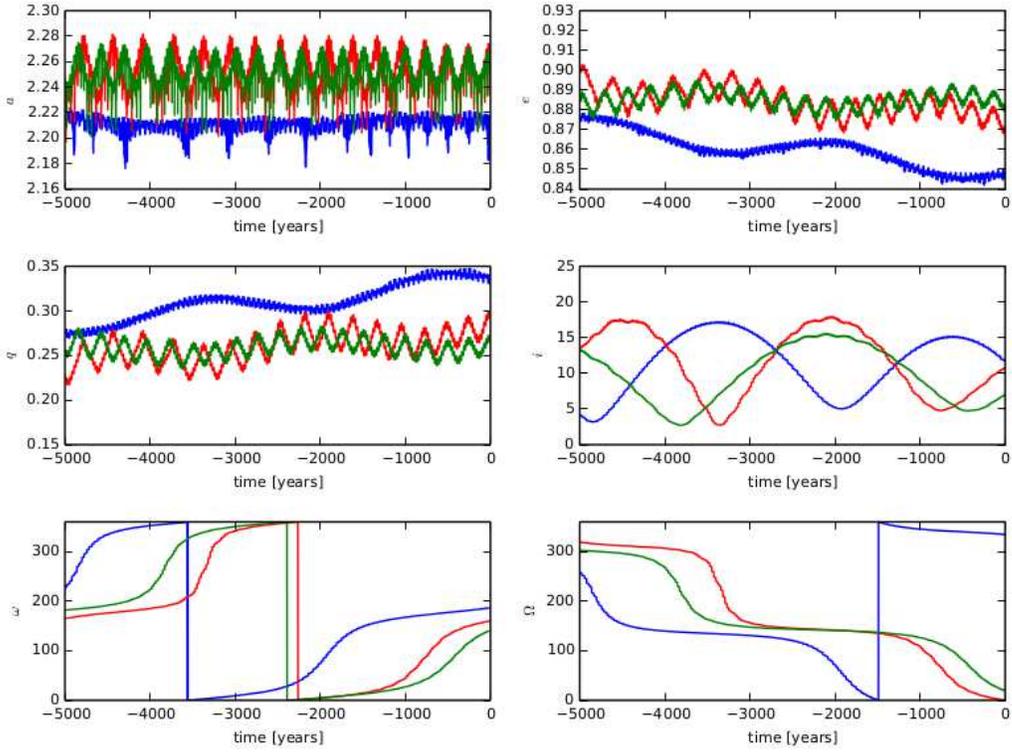}
\vspace{10.7cm}
\caption{The time evolution of orbital elements of 2005~UR (green), 2005~TF50 (red), and 2P/Encke (blue).}
\end{figure*}

\begin{figure*}
\centering
\includegraphics{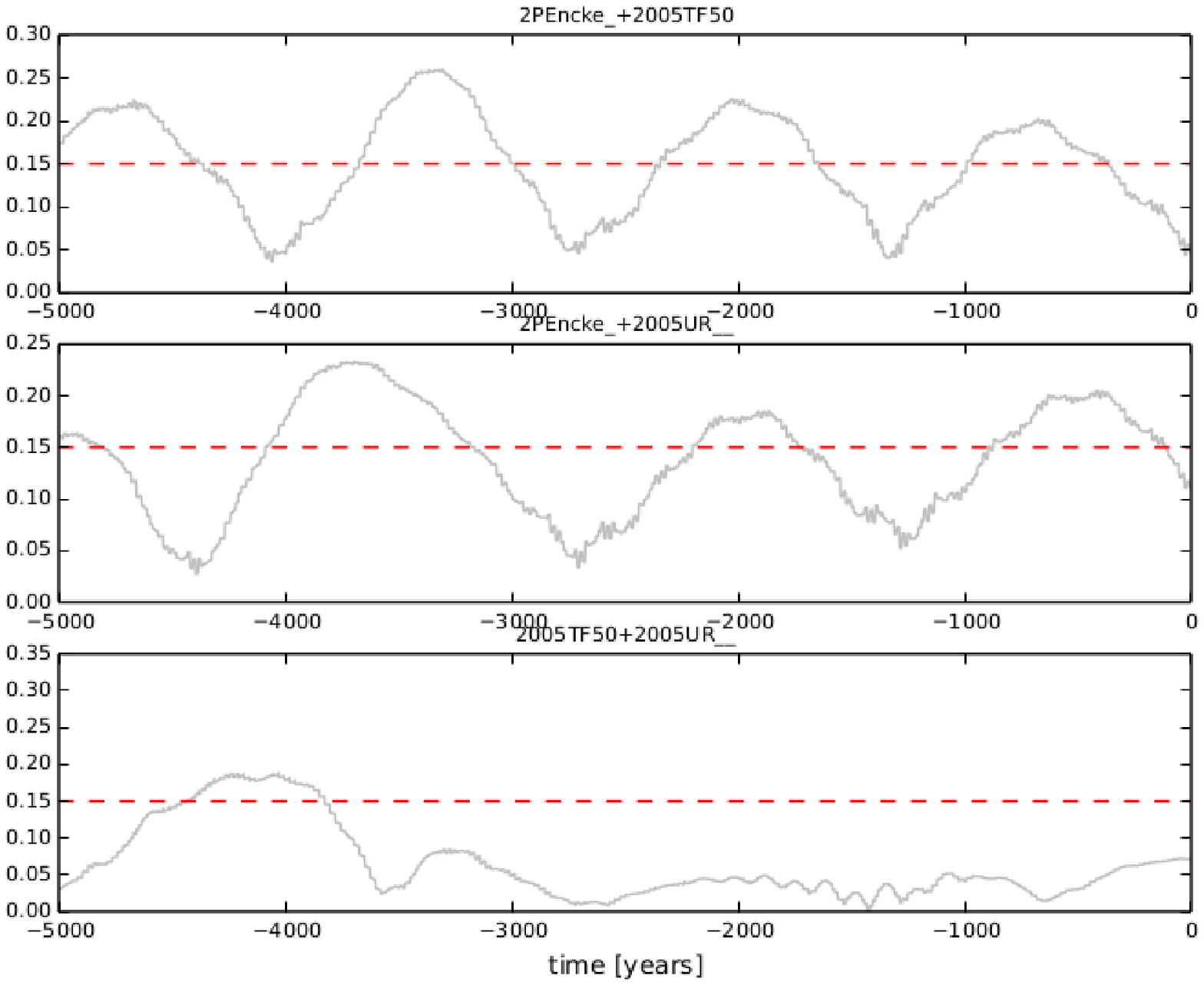}
\vspace{10.7cm}
\caption{Evolution of the $D_{S}$ criterion calculated by comparing the 
orbit of NEAs and the comet. The red dashed line shows the threshold value ($D_c=$0.15).}
\end{figure*}

\subsection{Modeling}

A numerical integration of the orbital parameters backwards in time has
been performed in order to test the link between the fireballs PF311015a
Okonek and PF311015b Ostrowite, two NEOs: 2005 UR, 2005 TF50 and comet
2P/Encke. For the integrations of the asteroids and test particles
representing fireball, the RADAU integrator in the Mercury software was
used (Chambers 1999). The test particles means a series of clones of the
radiant position and geocentric velocity of a fireball generated within
the measurement uncertainties, that were later converted into orbital
elements and propagated in the backward integration together with orbits
of NEOs. We generated 100 massless clones of the fireballs individually.

The model of the Solar System used in integrations
included: 8 planets, four asteroids (Ceres, Pallas, Vesta, and Hygiea),
and the Moon as a separate body. Additionally, we included the radiation
pressure here as well. The positions and velocities of the perturbing
planets and the Moon were taken from the DE406 (Standish 1998). The
initial orbital elements of asteroids 2005 UR and 2005 TF50 and comet 2P/Encke were taken
from JPL solar system dynamics web
site.\footnote{http://ssd.jpl.nasa.gov} Together with initial orbital
elements of asteroids and comet, the test particles were integrated to
the same epoch of the beginning of the integration. Next, the backward
integration was continued for 5000 yr.

During the evolution the generated stream has been widely dispersed in
longitude, therefore, we used Steel et al. (1991) criterion, $D_S$,
instead of a conventional similarity functions (Southworth \& Hawkins
1963, Drummond 1981, Jopek 1993). With the values being less than 0.15,
the evolution of the $D_S$ criterion reveals a link between Okonek
fireball and 2P/Encke, 2005~TF50, and 2005~UR, through 2300, 1600, and
1600 years, respectively (left panels in Figure 8). In case of Ostrowite
fireball, the evolution of the $D_S$ criterion shows similarity through
shorter period of times: 2000, 450, and 400 years with 2P/Encke,
2005~TF50, and 2005~UR, respectively (right panels of Figure 8).

If there is a link between two bodies then the value of the
dissimilarity criterion is very low at the moment of their separation,
and increases with time. In theory, analysing results of the backward
integration, we start in a moment when some time passed since the
separation. Therefore, we start with a higher value of the dissimilarity
criterion, then it decreases reaching a minimum value (at the possible
moment of the separation). And then it increases again because in the
integration two bodies are still treated as separate objects, as if the
separation did not occur -- unless we tell the program to stop
integration when a given condition is fulfilled. We would see more
complex image when involved in a study are objects which undergo a
stronger perturbation and are in resonance with a planet (particularly
Jupiter). The distance of an asteroid from the Jupiter's orbit,
characterised by the semimajor axis and aphelion distance, influences
the amplitudes and rates of changes of the perihelion distance ($q$),
eccentricity ($e$), and inclination ($i$). Moreover, all of used by us
in the study asteroids are close to 7:2 resonance with Jupiter. All of
this has its reflection is amplitudes and rates of changes of the
dissimilarity criterion as well. Thus, instead of a stable, linear
decreasing in going back in time to the possible separation moment we
observe sinusoidal curve.

Additional outcome of our work concerns the relation between asteroids
themselves. Figure 9 shows the time evolution of semimajor axis ($a$),
eccentricity ($e$), perihelion distance ($q$), inclination ($i$),
argument of perihelion ($\omega$), and longitude of ascending node
($\Omega$) of NEAs and the comet. Our results show orbital similarity
between 2005~TF50 and 2005~UR in the interval of almost 4000
years applying both $D_S$ and $D_{SH}$ (see Figure 10 for $D_S$ plot). The
lower values are obtained around 2600 years, which corresponds with low
values of $D_{S}$ and $D_{SH}$ when comparing asteroids' orbit with
orbit of 2P/Encke. This may suggest that around that time separation of
both asteroids might have occurred.

We generated 100 clones of asteroids 2005~UR and 2005~TF50, using their
orbital covariance matrix taken from the JPL Horizon. Analysing results
of the backward integration of those objects and their clones shows, as
expected, that an orbit calculated from a short data-arc span (6 days
for 2005~UR) would produce orbits with higher dispersion in time than
for a longer data-arc (26 days for 2005~TF50). However, the amplitudes
and rates of changes of orbital elements of clones have similar range
and pattern, especially for 2005~TF50, for which orbital uncertainties
of their nominal orbit are smaller.

The current and past $D_S$ values for each NEO and 2P/Encke are
not extremely low indicating that real separation of all these three
bodies might took place 20000-30000 years ago in one catastrophic event
which created the whole Taurid complex (Asher et al. 1993, Babadzhanov
et al. 2008). Leaving for a while a connection with 2P/Encke comet, we
can speculate about both NEOs and both fireballs origin. The orbits of
2005 UR and 2005 TF50 are very similar through a period of last 3600
years. Almost zero values of $D_S$ criterion are observed at moments of
$-700$, $-1300 \div -1500$ and $-2700$ years (Figure 10). What is
interesting is that the deep minimum of $D_S$ around the moment of
$-1500$ years was obtained for each NEO-fireball orbit combination (see
Figure 8). This epoch might be suspected as the time when larger body
was disrupted creating both NEOs and meteoroids which caused the
fireballs. Still we have take into account earlier minimum observed at
epoch around $-400$ years. This is the first deep minimum of $D_S$
observed for all NEO-fireball combinations. In case of the disruption
at that moment further backward integrations for earlier epochs have
no physical sense.

\section{Summary}

In this paper we presented an analysis of the multi-station observations
of two bright Southern Taurid fireballs which occurred over Poland on
2015 October 31. Moreover, we investigated their connection with two NEOs
and comet 2P/Encke. Our main conclusions are as follows:

\begin{itemize}

\item both meteors are similar with many aspects including brightness
higher than Full Moon, shape of the light curve, entry velocity,
persistent train and orbital parameters ($D_D$ of only 0.011), 

\item among over dozen of NEOs identified as possible parent bodies of
Taurid complex two, namely 2005 UR and 2005 TF50, have orbits which are
very similar to the orbits of observed fireballs (with $D_D<0.045$),

\item similarity of orbits of both fireballs and 2005 UR asteroid is
especially interesting due to the fact that the close flyby of this NEO
was observed exactly during last high maximum of Taurid complex shower
in 2005,

\item the numerical backward integration of the orbital parameters of
both fireballs and NEOs backwards in time, which has been performed in
this work, indicates many similarities between orbits of these objects
during past 5000 years. However, about 1500 years ago, $D_S$ criterion
has close to zero values for each of NEO-fireball, NEO-NEO and
fireball-fireball pair suggesting at that moment a disruption of a
larger body might took place, 

\item although, we could not have confirmed unequivocally the relation
between fireballs and 2005 UR and 2005 TF50, we showed that at least
both asteroids could be associated, having the same origin in a
disruption process that separates them.

\end{itemize}

The Taurid complex is certainly one of the most interesting objects in
the Solar System. It is able to produce both impressive meteor maxima and
extremely bright fireballs (Dubietis \& Arlt 2006, Spurn\'y 1994).
Additionally, it can be connected with catastrophic events like Tunguska
(Kresak 1978, Hartung 1993) and can affect the climate on Earth (Asher
\& Clube 1997). Accurate observations and analysis of all kind of bodies
associated with the Taurid complex are then very a important task,
demanding to continue and affecting the safety of our planet.

\section*{Acknowledgments}

This work was supported by the NCN grant number 2013/09/B/ST9/02168.

\end{document}